\documentclass[11pt]{article}
\usepackage{moriond-old,epsfig}
\begin{document}

\bibliographystyle{unsrt}    

\def\Journal#1#2#3#4{{#1} {\bf #2}, #3 (#4)}

\def\NCA{\em Nuovo Cimento}
\def\NIM{\em Nucl. Instrum. Methods}
\def\NIMA{{\em Nucl. Instrum. Methods} A}
\def\NPB{{\em Nucl. Phys.} B}
\def\PLB{{\em Phys. Lett.}  B}
\def\PRL{\em Phys. Rev. Lett.}
\def\PRD{{\em Phys. Rev.} D}
\def\ZPC{{\em Z. Phys.} C}

\def\st{\scriptstyle}
\def\sst{\scriptscriptstyle}
\def\mco{\multicolumn}
\def\epp{\epsilon^{\prime}}
\def\vep{\varepsilon}
\def\ra{\rightarrow}
\def\ppg{\pi^+\pi^-\gamma}
\def\vp{{\bf p}}
\def\ko{K^0}
\def\kb{\bar{K^0}}
\def\al{\alpha}
\def\ab{\bar{\alpha}}
\def\be{\begin{equation}}
\def\ee{\end{equation}}
\def\bea{\begin{eqnarray}}
\def\eea{\end{eqnarray}}
\def\CPbar{\hbox{{\rm CP}\hskip-1.80em{/}}}

\vspace*{4cm}

\title{ Intensity Mapping with the 21-cm and Lyman Alpha Lines}
\author{ J. B. Peterson and Enrique Suarez }

\address{Department of Physics, Carnegie Mellon University, 5000 Forbes Avenue,\\
Pittsburgh PA 15217, USA}

\maketitle\abstracts{The 21-cm and Lyman Alpha lines are the dominant line-emission spectral features at opposite ends of the
spectrum of hydrogen. Each line can be used to create three dimensional intensity maps of large scale structure. The sky brightness at low redshift due to Lyman Alpha emission 
is estimated to be 0.4 Jy/Steradian, which is brighter than the zodiacal light foreground.
}

\section{Introduction}

\indent This paper is about foregoing individual galaxy detections and, instead, directly mapping cosmic structure using the 21-cm and Lyman alpha (Ly$_\alpha$) lines.  For almost a century the {\it galaxy} has been the glowing test particle used to trace the cosmic expansion. A galaxy is a aggregation of as much as $\sim10^{12}$ solar masses of dark matter and gas, concentrated to a density millions of times the cosmic average. This concentration stimulates star formation and the resulting increment of specific intensity on the sky can be detected with optical telescopes at redshifts that now approach eight. Although galaxies will continue to be of great value in tracing cosmic structure, we examine an alternative: using line emission of widely distributed gas as the tracer of cosmic density structure.

Previously, all studies of three dimensional large scale (cosmic-web) structure have used bright galaxies to carry out redshift surveys. In this type of study, individual galaxies are selected from images of the sky, a set of slit or fiber masks are fabricated, spectra are recorded, and each spectrum is examined. If spectral features are detected with sufficient signal to noise, usually about 5, an individual entry is made in a redshift catalog. Modern redshift surveys collect 100,000 or more redshift entries. The study of large scale structure then proceeds by examining the catalog. This is, clearly, a very laborious process, but also an insensitive one. Redshift surveys miss a great deal of information by recording spectra only on a sparse set of site lines and, in addition, the five sigma threshold means many spectra with substantial signal are discarded. In contrast, intensity mapping is a technique that requires no detection threshold, and therefore uses all the available spectral information on each line of sight. This technique also rapidly covers sky by using large beams, or pixels, leading to orders of magnitude improvement of mapping speed compared to traditional galaxy-based survey techniques.

The term Intensity Mapping (IM) is short for `three dimensional mapping of the specific intensity due to line emission. To use this technique, one assumes the spectral density of the flux received along each line of sight is due to a single emission line (or a template of multiple lines), and the emissivity of the line is linearly related to mass density. Then, the received spectral density can be translated directly to cosmic density just by translating the received frequency to comoving distance, and the brightness to mass density. 

One of the main advantages of intensity maps is that they need only 1-10 arc-minute angular resolution, which allows for rapid coverage of large comoving volumes. Such coarse resolution is allowed because much of cosmology focuses on the study of linearly evolving structure. The smallest linear scale is $\sim 10$ Mpc which subtends a few arcminutes. Assuming a typical average spacing of galaxies around 1 Mpc, voxels resolving 10 Mpc would on average contain $\sim$1000 galaxies. Traditional redshift surveys would record spectra for only the brightest of these galaxies. An intensity mapping survey, on the other hand, includes the emission from all the galaxies.

\begin{figure}[htbp]
\begin{center}
\includegraphics[width= 5.0in,trim=30 0 0 30]{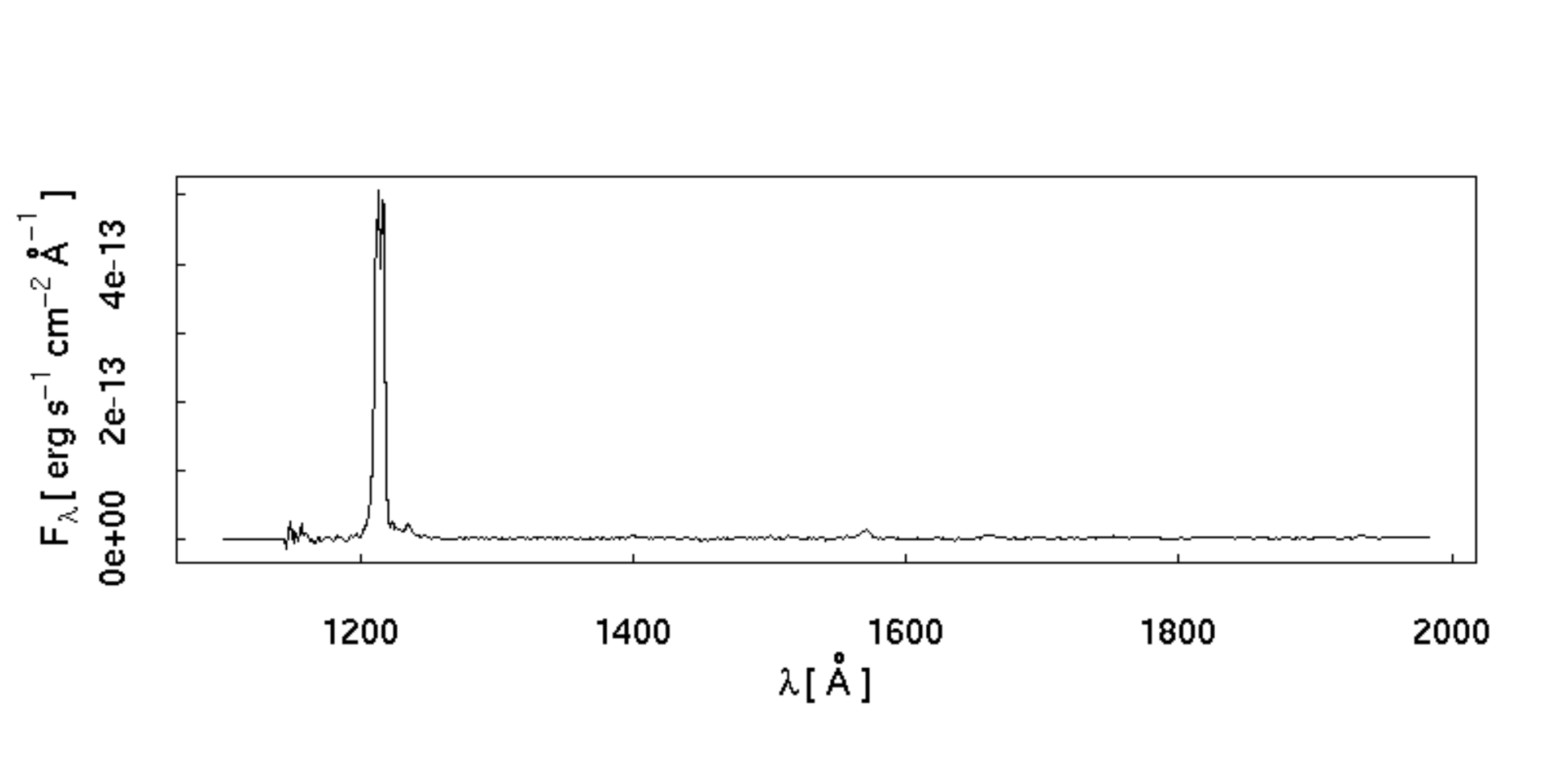}
\end{center}
\caption{\scriptsize{9} {\bf The line spectrum of the galaxy NGC 0262, at z $\sim$ 0.015.}
The Ly$_\alpha$ emission line is very prominent and several orders of magnitude brighter than the small bumps in the spectrum due to the OVI doublet and CIV line. Image obtained from NASA/IPAC Extragalactic Database (NED)}
\label{fig:Spectrum NGC 0262}
\end{figure}

As mentioned above, intensity mapping assumes that the spectral density of the flux received along each line of sight is due to a single emission line. In the spectrum of a typical spiral galaxy two lines stand out: both the 21-cm line and the Ly$_\alpha$ line (see Figure \ref{fig:Spectrum NGC 0262}) are much brighter than neighboring continuum emission, and are well separated from other spectral lines. That means that each line can be used to directly create three dimensional maps of cosmic structure, via the intensity mapping technique. 

\section{21-cm Intensity Mapping}

Intensity mapping is a natural extension of previous work in radio astronomy. Throughout the history of radio astronomy, instruments have been optimized either for brightness sensitivity (single dishes) or resolution (dilute aperture synthesis arrays), but not both. This limitation was due to the lack of signal processing computational power needed to allow focal plane or synthesis arrays with more than about 50 elements. Now, with the Moore's Law reduction of processing cost, 512 element systems are under construction, and 10,000 element arrays will be possible within a few years. This increased signal-processing bandwidth makes high-sensitivity 21-cm intensity mapping possible.

Several authors ~\cite{battye04} \cite{sethi05} \cite{wyithe08} \cite{chang08} \cite{peterson09} have developed the idea of 21-cm intensity mapping at redshifts below six. At these redshifts, cosmic UV flux is high enough to keep most of the volume of intergalactic space ionized. Then, neutral gas can only survive in dense self shielded regions such as spiral galaxies. Thus, at low redshifts, with gas localized to galaxies, it is tempting to pursue the traditional redshift survey using the 21-cm line. Research teams using the Parkes and Arecibo telescopes have carried out surveys that detect the galaxies individually. Nevertheless, despite years of telescope time at large telescopes, these surveys do not extend beyond redshift $z$ $\geq$0.1. One reason for this redshift limit is the 21-cm flux from an individual galaxy declines as $z^{-2}[1+z]^{-2}$.  In contrast, the sky brightness contrast due to the 21-cm line declines only as $[1+z]^{-1/2}$, making intensity mapping the productive option as the redshift increases. At higher redshift, before reionization at $z \sim 10$, neutral gas clouds extend to sizes exceeding 1 Mpc, and 21-cm intensity mapping is a natural choice. In fact, the world-wide suite of 21-cm Epoch-of-Reionization telescopes \cite{morales09} are all designed for intensity mapping.

Moving well beyond the redshift limit of traditional redshift surveys, Chang $et al$. \cite{chang10} have used intensity mapping to detect emission of the 21-cm line at redshift $z$ $= 0.8$, using the Green Bank Telescope. By stacking 21-cm emission from the regions surrounding 5000 bright galaxies in the optical DEEP2 redshift survey, they detected the 21-cm signal in just 12 hours of telescope time. An important part of this accomplishment is detection of 21-cm sky structure five orders of magnitude below the synchrotron glow of the Galaxy. 

While 21-cm intensity mapping seems well on its way to becoming widely accepted, there are other spectral lines from hydrogen that are currently being used to detect galaxies at high redshift, and these lines might be useful for directly mapping linearly evolving structure. For example, in the ultra-violet end of the spectrum, Lyman Alpha (Ly$_\alpha$; $\lambda_{e}$ $=1215$ $\AA$), has not yet been used for intensity mapping. We will discuss below why it is also a worthy candidate.

\begin{figure*}

\begin{center}
\includegraphics[width=4.0in ]{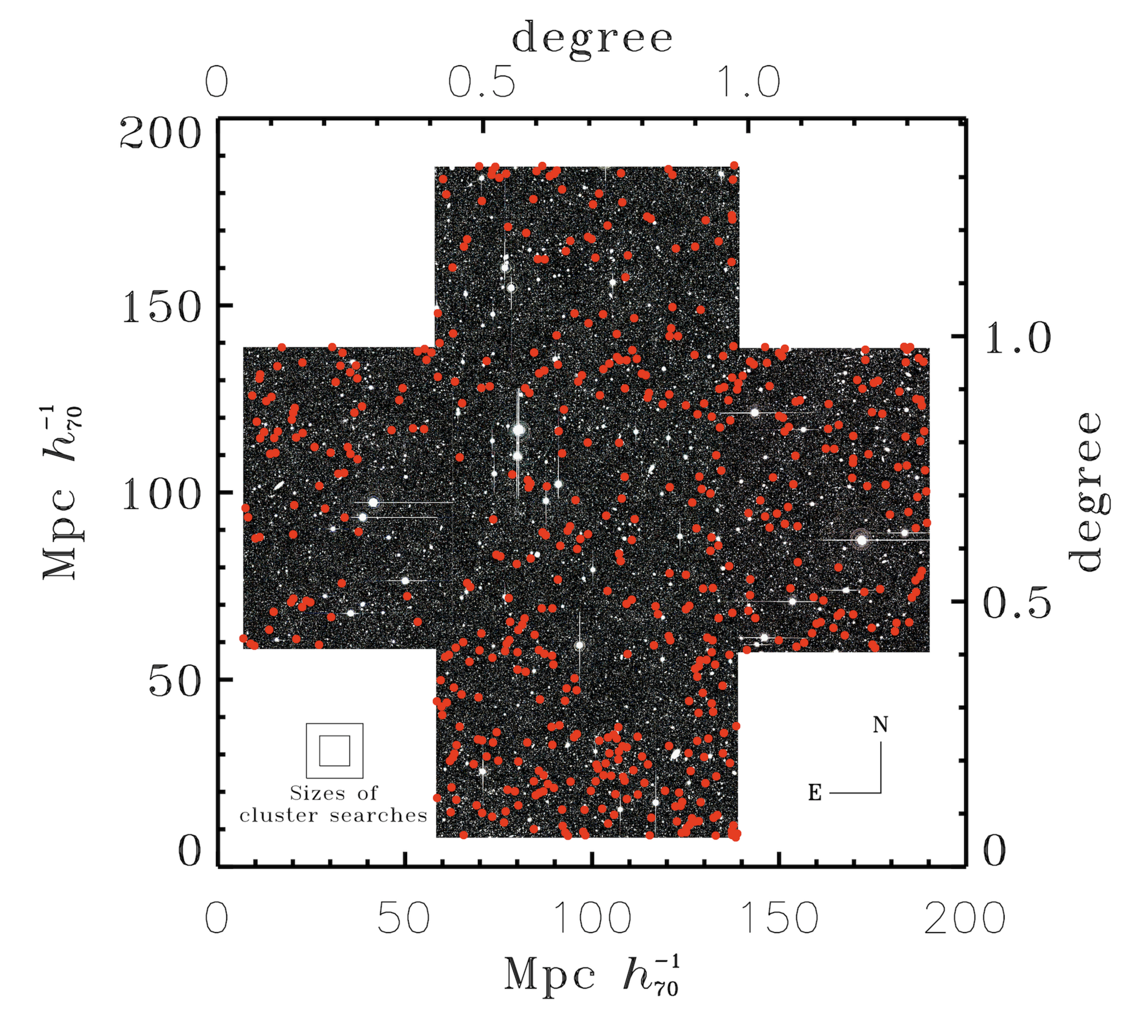}
\end{center}
\caption{\scriptsize{9} {\bf Lyman Alpha Emitters at redshift $z$ $= 5.9$}   
Red dots show the locations of Lyman Alpha Emitter candidates detected in narrowband-wideband image ratios. Large Scale Structure is apparent in the form of clusters and voids in the LAE locations. Plot from Ouchi et $al.$ (2005) \protect \cite{ouchi05}, adapted by Kevin Bandura}
\label{fig:LAE-sub}
\end{figure*}

\section{Lyman Alpha Intensity Mapping}

Figure \ref{fig:LAE-sub} shows the positions of over one hundred Lyman Alpha Emitter (LAE) candidates detected in deep mosaic observations using Subaru/Suprimecam. These candidates were selected by finding five-sigma bright outliers in narrowband versus broadband flux. The narrow band filter was tuned to Ly$_\alpha$ redshifted by $z$ $= 5.9$. In follow-up of similar observations \cite{ouchi10}, over 200 candidates were later confirmed as Ly$_\alpha$-bright galaxies. Large scale structure is apparent in figure \ref{fig:LAE-sub}  (and confirmed statistically \cite{ouchi10}), since it contains several tight concentrations of sources and voids. The Subaru LAE image gives confidence that Ly$_\alpha$ intensity mapping, which amounts to taking many narrowband wide-field images simultaneously, can be used to rapidly map cosmic structure.

As we suggested above, intensity mapping is much more efficient than the narrowband technique used for traditional LAE surveys. Rather than restricting the spectral range with a single narrowband filter, which discards $\sim$ 99 \% of the observed spectrum, intensity mapping observations would collect a spectrum across a broad band for every pixel. The redshift slice shown by the red dots in Figure \ref{fig:LAE-sub} and hundreds of other slices could be imaged simultaneously in each exposure, allowing a factor $\sim$ 100 in speed increase. 

Unlike radio astronomy, optical/UV astronomy makes use of detector arrays with millions of simultaneously operating detectors. In addition, even small telescope apertures offer arc-second angular resolution. Therefore, astronomers operating at optical/UV wavelengths are used to detecting galaxies individually, and all of their equipment is designed to do so. Thus, the thought of abandoning individual galaxy detections in favor of improving mapping speed and bandwidth, may seem like a step backward. However, we argue below that the speed advantages are substantial and, by calibrating the IM technique using redshift surveys of small fields, one can gain confidence that these types of surveys will successfully map cosmic structure.

An additional efficiency factor in favor of intensity mapping comes from the high-pass filtering used to make the individual LAE detections using the narrowband images such as figure \ref{fig:LAE-sub}. At these high redshifts, this high-pass filtering is likely discarding substantial flux since Ly$_\alpha$ photons tend to diffuse by resonantly scattering off of neutral hydrogen. Clear evidence of this scattering comes from the study of quasar spectra. By redshift $z$ $\sim$5.9, most quasars spectra show substantial Gunn-Peterson decrements. However, photons taken out of the line of sight in the Gunn-Peterson effect are typically not absorbed, but are simply scattered to other lines of sight and surely present in the form of low surface-brightness extended halos around bright sources. Such halos would be discarded by high-pass filtering, and therefore would often be missed. However, in a few images extended Ly$_\alpha$ emission has recently been noticed and described as Lyman Alpha Blobs \cite{matsuda04} \cite{steidel00}. 

Diffuse emission is a subject where simulations are more advanced than observations. Zheng $etal$. \cite{zheng11} estimate that the random walk distances for Ly$_\alpha$ photons can be as large as 1 Mpc before the photons redshift out of the resonance. These simulations indicate that ratio of photon counts in the diffuse (arcminute-sized) halo compared to central (arcsecond-sized) LAE galaxy may be as high as 100. If this is correct, intensity mapping could capture 100 photons for each that survives the high pass filter of the traditional survey.  

Returning to the Subaru image, figure  \ref{fig:LAE-sub}, the Ly$_\alpha$  bright galaxies that meet the five sigma selection criterion to produce a red dot are in the high luminosity region of the luminosity function. It is important to remember that for each red dot there are surely hundreds of nearby dimmer LAE sources. The red dots are like tips of icebergs, each marking the position of a great mass of unseen sources.

\begin{figure*}
\begin{center}
\includegraphics[width = 3 in]{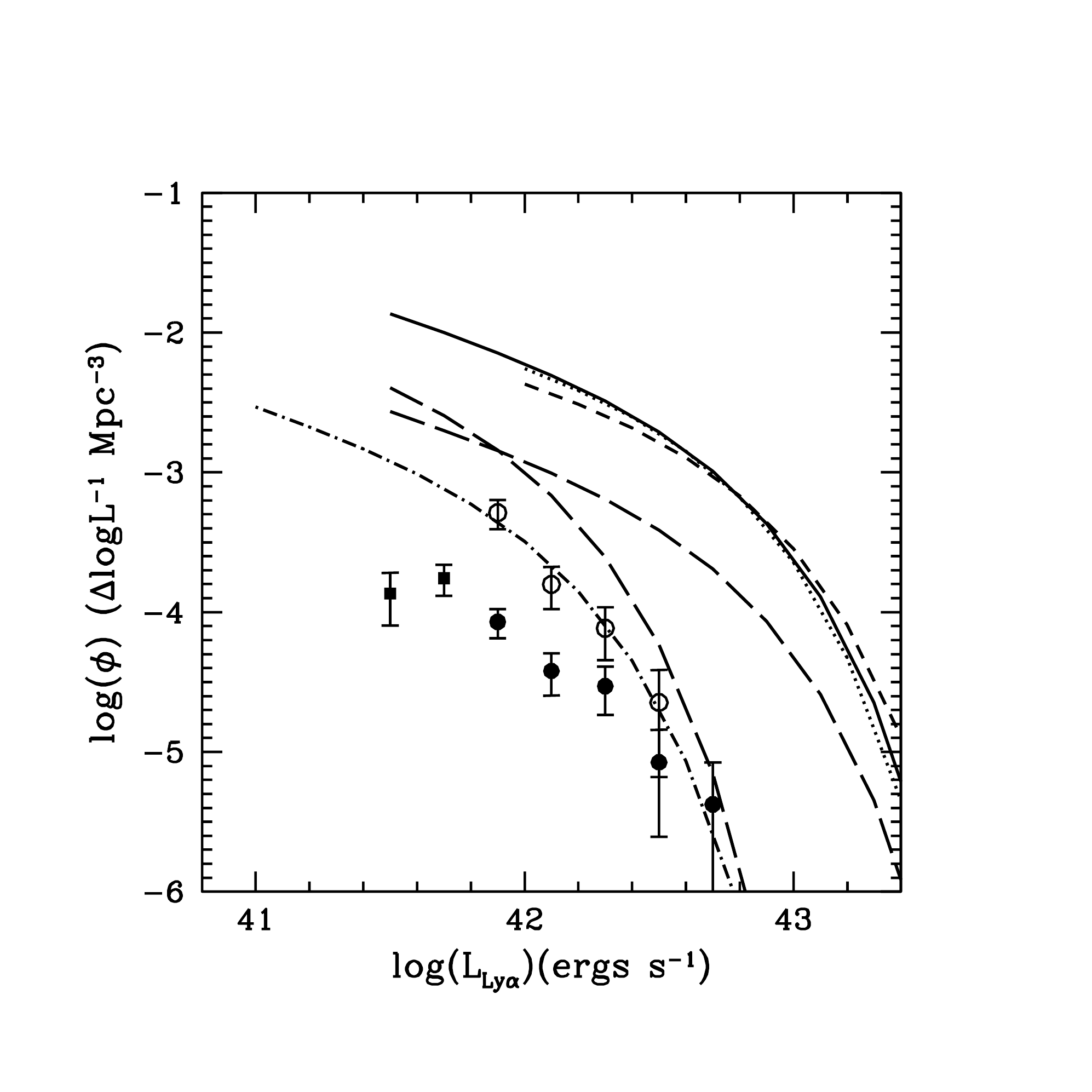}
\end{center}
\caption{\scriptsize{9} {\bf Ly$_\alpha$ luminosity functions.}  Space density of Ly$_\alpha$ emitting galaxies (0.2 $<$ $z$ $<$ 0.35) per logL$_{Ly\alpha}$ both as measured ($filled$ $circles$, all five fields; $filled$ $squares$, CDFS, GROTH, and NGPDWS fields) and with an evaluation accounting for incompleteness ($open$ $circles$). The lines represent comparisons with Ly$_\alpha$ LFs at high redshifts. $Solid$ $line$: van Breukelen et al. (2005) \protect \cite{vanB05} at 2.3 $<$ $z$ $<$ 4.6. $Dotted$ $line$: Gronwall et al. (2007) \protect \cite{gronwall07} at $z$ $\sim$ 3.1. $Short-dashed$ $line$: Ouchi et al. (2008) \protect \cite{ouchi08} at $z$ $\sim$ 3.1. The $dot-dashed$ LF is derived from a least-squares fit on the 5 brightest points. The $long-dashed$ lines show the impact of a factor of 5 decrease of L$^{*}$ (the nearest curve to the data points) or $\Phi^{*}$ in the LF of van Breukelen et al. (2005); this factor corresponds to the decrease of the UV LD from $z$ $\sim$ 3 to $z$ $\sim$ 0.3. Figure and caption taken from Deharveng, J-M. et al. (2008) \protect \cite{deharveng08}, which analyzed data from GALEX}
\label{fig:LF-LAE}
\end{figure*}

We have described above several types of mapping-speed efficiency factor that favor intensity mapping versus the narrowband imaging technique. We now collect these to attempt a rough estimate of the total mapping-speed advantage. Counting orders of magnitude, for redshift $z$ $\sim$ 6: a) recording spectra versus narrowband filtering--2, b) the use LAE flux from sources below the five sigma detection threshold--1, c) the use of the photons in resonant scattering halos--1-2. d) the use of arc-minute size pixels--2. Intensity mapping  could therefore be six to seven orders of magnitude more efficient than the narrowband imaging technique. This means that expensive 10 meter class instruments like Subaru are not needed for an intensity mapping survey. Much smaller apertures can be considered.

The required aperture for a Ly$_\alpha$ intensity mapping telescope could be less than one meter, and this is small enough that the telescope can be placed on balloon-borne platforms or spacecraft. This opens the possibility that redshift ranges that are largely absorbed by the atmosphere, 0 $\leq$ $z$ $\leq$ 3, and  $z$ $>$ 10 could be observed.  At the low redshift end, an important target of an intensity mapping survey would be the detection of Baryon Acoustic Oscillation features.  However, note that, at low redshift, LAEs are typically dimmer than at high redshift, and the resonant scattering halos are likely absent at low redshift. In turn, this would make the observations more challenging than at redshifts $z$ $\geq$6.

While we know of no research group that has so far  conducted an intensity mapping Ly$_\alpha$ survey, several teams have estimated the luminosity functions (LF) of Lyman Alpha Emitters at various redshifts. For example, Deharveng $et$ $al$. (2008) \cite{deharveng08} constructed Figure \ref{fig:LF-LAE}, which compares  their estimate of the LF of LAEs to those of other researchers. The low-redshift measurements were made using the GALEX observatory. At higher redshift, the detections were made using ground-based telescopes. In either case, as described above, spatial and spectral high pass filters are used, and only the bright end of the luminosity function is detected. Unfortunately, then, the bulk of Ly$_\alpha$ emission may be missing from these luminosity functions. However, these LFs are the best measurements available of Ly$_\alpha$ source strengths, so we proceed to use them to calculate the sky brightness due to Ly$_\alpha$.

\begin{figure}[!ht]
\begin{center}
\includegraphics[scale=1.25,height=3in,trim=30 0 0 0]{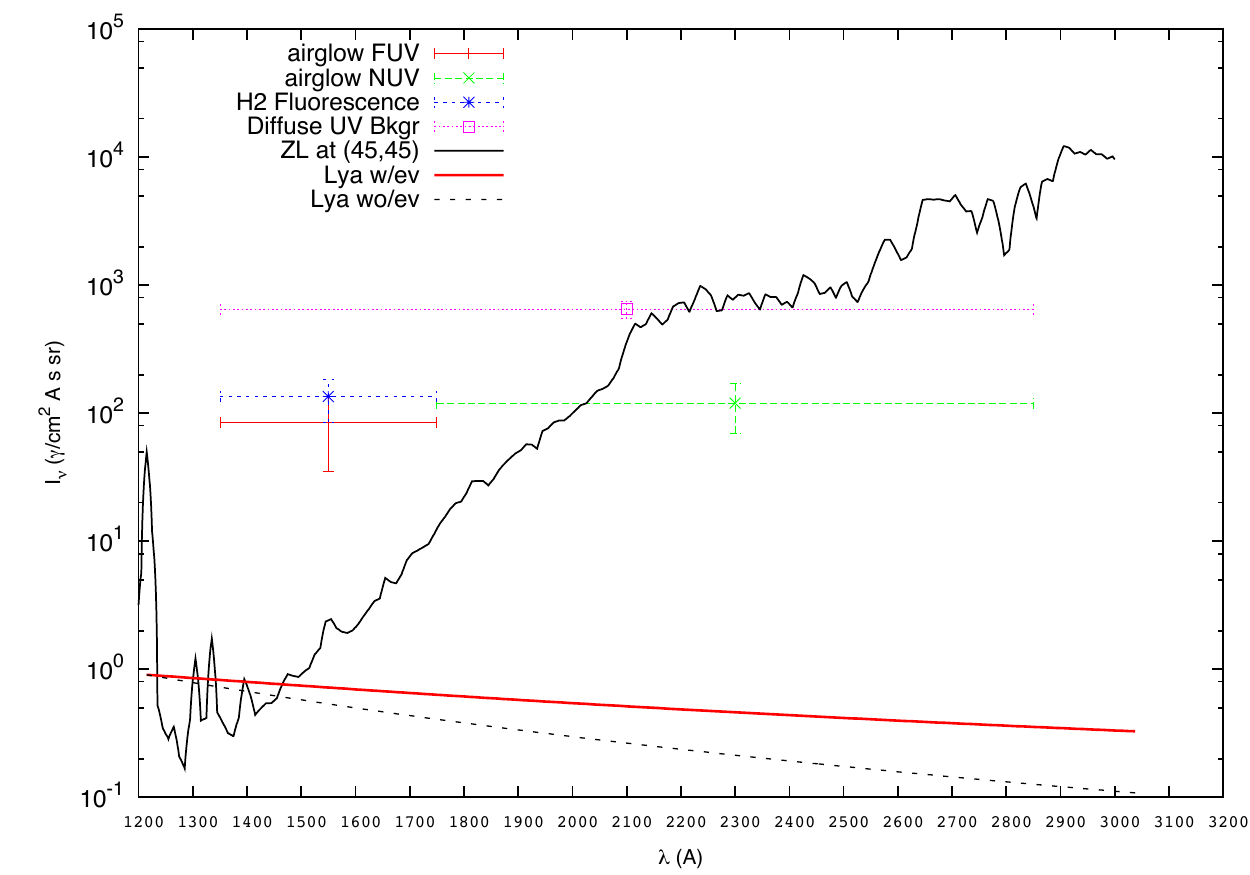}
\caption{\scriptsize{9} {\bf The Ly$_\alpha$ Sky Brightness.}  The red curve show the estimated brightness due to Ly$_\alpha$ emission. Here modest luminosity function evolution has been assumed.  The dashed black curve assumes no evolution. Additional sources of sky brightness include: Ariglow, H$_2$ fluorescence, Diffuse UV  \protect \cite{sujatha09}; the plotted values for these sources represent the maximum amount of emission detected by GALEX. Zodiacal light \protect \cite{leinert98}, is estimated for  the direction (45$^{\circ}$, 45$^{\circ}$) in ecliptic coordinates, measured from the sun. \label{fig:LySB}}
\end{center}
\end{figure}

\section{Ly$_\alpha$ Sky brightness}
\indent Integrating the GALEX luminosity function yeilds\begin{equation}
\frac{dP}{dV} = 10^{39} \ erg \ s^{-1} \ Mpc^{-3}
\end{equation}
which translated to Ly$_\alpha$ brightness 
\begin{equation}
I_\nu = \left( {4 \times 10^{-27}}\right) \frac{1}{\left( \Omega_\Lambda+ \Omega_m \left( 1+z \right)^{3} \right)^{1/2}} \ W \ m^{-2} \ sr^{-1} 
\end{equation}

The Ly$_\alpha$  brightness is compared to other sources of sky brightness in figure \ref{fig:LySB}. The zodiacal light (ZL) in the UV region is due to the sunlight scattered by interplanetary dust. The ZL brightness is a smooth function angle of observation with respect to the sun, and the ZL spectrum closely matches the solar spectrum. We estimate the ZL brightness using the solar spectrum of Leinert et $al$. (1997) \cite{leinert98}. At low redshift, the brightness due to Ly$_\alpha$ actually exceeds the (ZL) foreground glow. As redshift increases, the zodiacal light exceeds the Ly$_\alpha$ glow, but the ZL is spatially very smooth and can likely be filtered or modeled out very precisely. Additionally, we also have to worry about the diffuse UV background emission, which in both $GALEX$ bands, FUV and NUV, was observed \cite{sujatha09} to be 500 - 800 $\gamma \ cm^{-2} \  s^{-1} \AA^{-1} sr^{-1}$. The source of this emission is the subject of controversy, and some of it may be due to Ly$_\alpha$. We also plot the upper limit to the H$_2$ fluorescence emission, from data collected by $GALEX$ which constrain the flux to the range 0 - 250 $\gamma \ cm^{-2} \  s^{-1} \AA^{-1} sr^{-1}$ in the FUV, with no contribution in the NUV \cite{sujatha09}. For balloon-borne telescopes or those in low earth orbit another competing source is Airglow, which is produced in the Ionosphere's E and F layers, at altitudes of $\sim$ 90 km, and above 150 km, respectively, as well as a contribution from the Geocorona. Observations by $GALEX$ indicate Airglow emission to be 85 $\gamma \ cm^{-2} \  s^{-1} \AA^{-1} sr^{-1}$ \cite{sujatha09}. These are all the competing foregrounds we know of and the foreground brightness rises to exceed the Ly$_\alpha$ brightness by $10^4$ at the high end of the plotted redshift range. This compares well to 21-cm, where the galactic synchrotron brightness is a factor $\sim 10^4$ brighter than the 21-cm brightness. The foregrounds should be easier to subtract for Ly$_\alpha$, since ZL is very smooth on the sky and sun-synchronous.

The Ly$_\alpha$ brightness structure can be estimated by assuming the Ly$_\alpha$ emission traces density. The usual bias factor $b ={{ \delta I/I} \over {\delta \rho /\rho}}$ will be needed to translate brightness structure to density structure and estimates of $b$ will no doubt be the subject of lengthy discussion.

\begin{figure}[!ht]
\begin{center}
\includegraphics[scale=1.25,height=3in,trim=30 0 0 0]{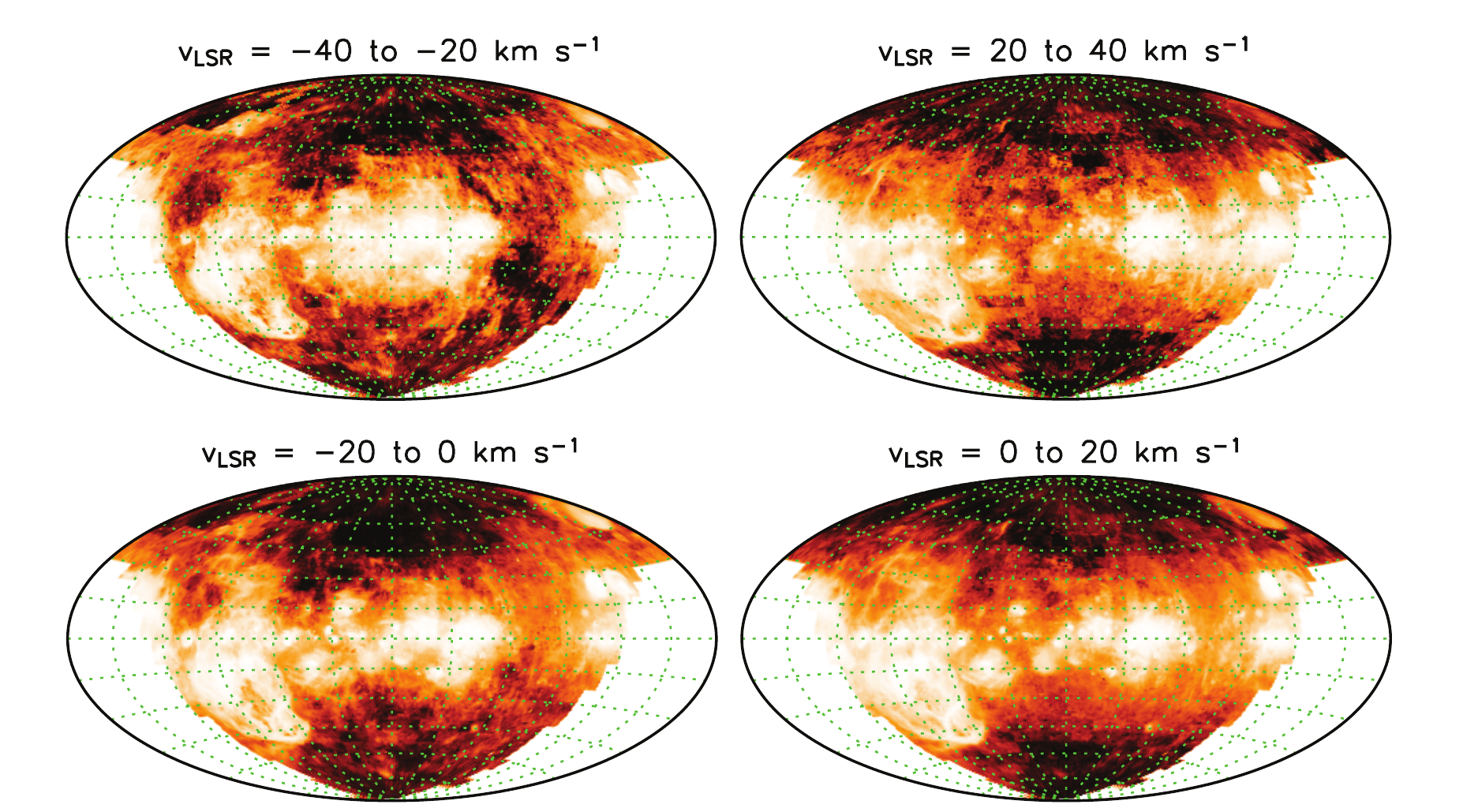}
\caption{\scriptsize{9} {\bf Intensity maps of the $H_\alpha$ emission in the Milky Way} Images made using the Wisconsin H-alpha Mapper, which uses a 0.6 m aperture. From Matsuda et $al$. (2004) \protect \cite{matsuda04}}
\label{fig:hamaps}
\end{center}
\end{figure}

\section{Techniques for Optical/UV Intensity Mapping.}  

While cosmological intensity mapping in the UV/optical has not been attempted, optical intensity maps of the Milky Way have been made using the Wisconsin H-alpha Mapper (WHAM). This system uses a tunable Fabre-Perot narrowband filter to select $H_{\alpha}$ emission at a variety of  doppler shifts within the Galactic velocity profile. Images from WHAM \cite{hafner03} are shown in Figure \ref{fig:hamaps}. The telescope aperture is only 0.6 m, but by using a wide field, three degrees, the instrument can rapidly map the volume brightness of the line. These images are rather encouraging, given that they show what is possible with a very small telescope specifically designed for intensity mapping. Furthermore, the images shown have angular resolution three arcminutes, similar to that needed for the study of cosmological large scale structure. The choice of a tunable narrow filter makes sense for this instrument since it is designed primarily to observe in the rather narrow velocity range of the Galaxy. For a survey of cosmological large scale structure, one would not want a narrow filter but instead would like to cover a much wider frequency range in each exposure.

For redshifts 0 $\leq$ $z$ $\leq$ 3, Ly$_\alpha$ observations must be carried out above the atmosphere, but at these short wavelengths optical elements have substantial attenuation. Optical system must be very simple at these wavelengths. One example of an instrument that may be capable of making Ly$_\alpha$ intensity maps is the GALEX spacecraft instrument, which is equipped with a slit-less spectroscopy mode using a Grism. It may be possible to use the GALEX grism, but also rotate the spacecraft through a set of paralactic angles. Each image is a projection of the intensity map, with the projection direction precessing along a cone. An inverse Radon transform of this set of observations should to allow tomography, essentially a CAT-scan of cosmic structure. 

At optical wavelengths, Ly$_\alpha$ redshifts 3 $\leq$ $z$ $\leq$ 7, very high mapping speed can be achieved using an integral field spectrometer. Here, each lens of a microlens lens array defines a large pixel on the sky and collects the light from that pixel into an optical fiber. These fibers are then routed to populate the slit of a spectrometer allowing a fast wide field, three dimensional observation. 

For redshift 7 $\leq$ $z$ $\leq$ 9, the late stages of reionization, Sliva $et$ $al$. \cite{silva12} have thoroughly studied Ly$_\alpha$ IM and suggest that a sub-orbital or orbital instrument may be useful at these redshifts. Aiming at redshift $z$ $\sim$15, The Low Resolution Spectrometer on the CIBER sounding-rocket instrument produces spectra of 1280 pixels in each observation. It is impressive that an instrument with a 5 cm aperture, observing for just 8 minutes in an intensity mapping mode, has the potential to detect cosmological large scale structure. At redshift $z$ $\sim$15 stars are likely shrouded in neutral gas, so the Lyman-limit edge may produce more sky structure than the Ly$_\alpha$ line.

\section{Line Confusion} A significant difference between 21-cm IM and Ly$_\alpha$ IM is the issue of line confusion. The 21cm line is the strong line in the emission spectrum of a galaxy with the {\it longest} wavelength. While there are Rydberg lines of longer wavelength, these are much weaker than 21-cm. Because there is no longer-wavelength line to cause confusion as the redshift increases, some authors contemplate 21-cm intensity mapping at redshifts as high as 50.  In contrast, Ly$_\alpha$ dominates the opposite extreme end of the spectrum of the hydrogen atom. Now, there is potential for confusion with other lines such as the rest of the Lyman series, H-alpha, and the lines of other species. However, the other Lyman series lines are at shorter wavelength, and are much weaker than Ly$_\alpha$, validating the assumption that a single line dominates sky brightness. In deep intensity maps, though, the other Lyman series lines may be detectable by cross correlation at the appropriate lag in the spectrum. In this case, rather than causing confusion, the rest of the Lyman series serves to confirm the Ly$_\alpha$ structure. At redshift $z$ $= 4.4$ Ly$_\alpha$ crosses the rest frame wavelength of H-alpha. Longward of this wavelength ($\lambda$ $= 656$ nm), the two intensity maps will overlap in the data cube. However, most of the LAE candidates in narrowband images seem to be Ly$_\alpha$ rather than H$_\alpha$ sources, so the assumption of Ly$_\alpha$ dominance should be accurate.

\end{document}